\documentclass[aps,twocolumn,showpacs,groupedaddress,nofootinbib]{revtex4-1}

\usepackage{graphicx}
\usepackage[utf8]{inputenc}
\usepackage{natbib,amssymb,graphicx}
\usepackage{pstricks} 

\usepackage[dvips]{hyperref}

\newcommand{\dd}{{\rm d}}

\newcommand{\RHO}{\mbox{$\rho^0$}}

\newcommand{\UNIT}[1]{\mbox{$\,{\rm #1}$}}

\newcommand{\GeV}{\UNIT{GeV}}

\newcommand{\fm}{\UNIT{fm}}

\newcommand{\ie}{i.e.}

\begin{document}

\title{\texorpdfstring{Color transparency in hadronic attenuation of
    $\rho^0$ mesons}{Color transparency in hadronic attenuation of
    rho0 mesons}}
\author{K.~Gallmeister} \author{M.~Kaskulov} \author{U.~Mosel}
\affiliation{Institut f\"ur Theoretische Physik, Universit\"at
  Giessen, Germany}

\begin{abstract}
  Ongoing experiments at JLAB investigate the nuclear transparency in
  exclusive $\RHO{}(770)$ electroproduction off nuclei.  In this work
  we present transport model predictions for the attenuation of
  \RHO{}s in nuclei and for color transparency (CT) effects as
  observable at CLAS with a 5~\GeV{} electron beam energy. A full
  event simulation developed here permits to study the impact of
  actual experimental acceptance conditions and kinematical cuts.  The
  exclusive $(e,e'\RHO{})$ cross section off nucleons is described by
  diffractive and color string breaking mechanisms extended toward the
  onset of the deep inelastic regime. Different hadronization and CT
  scenarios are compared. We show that a detailed analysis of
  elementary cross section, nuclear effects and experimental cuts is
  needed to reveal the early onset of $\rho$-CT at present JLAB
  energies.
\end{abstract}


\date{\today}

\maketitle


\section{Introduction}

Electroproduction experiments of hadrons on nuclear targets may offer
a chance to observe the effect of color transparency (CT), that is the
phenomenon of vanishing (or very small) final state interaction (FSI)
of the produced hadron with the surrounding nucleons, at high momentum
transfer~\cite{Brodsky:1988xz,Frankfurt:1992dx,Jain:1995dd,Farrar:1988me,Jennings:1991rw}.
The HERMES experiment at DESY~\cite{Airapetian:2002eh} (see
\cite{Aclander:2004zm,Carroll:1988rp,Adams:1994bw} for other experiments at higher energies) looked for such an effect in the electroproduction of \RHO{}
mesons on nuclei. While a rise of the transparency $T_A$\footnote{The
  nuclear transparency for a certain reaction process is usually
  defined as the ratio of the nuclear cross section per target nucleon
  to the one for a free nucleon, \ie{}~$T_A=\sigma_A/A\sigma_N$.}
with photon virtuality $Q^2$ was indeed observed and taken as evidence
of a $\rho$-CT effect other studies showed that the same data could be
well understood in a sophisticated transport
calculation~\cite{Falter:2002vr} or even a Glauber
calculation~\cite{Kopeliovich:2001xj}.  In these studies the observed
rise of $T_A$ was attributed to the change of the coherence length,
\ie{}~to the $Q^2$-dependence of shadowing in the entrance channel.
The observation of $\rho$-CT, therefore, demands a kinematical regime
where one is less sensitive to the resolved hadronic interactions of
the incoming photons.  A new experiment at JLAB has been designed to
work under kinematical conditions that keep the coherence length small
and nearly constant~\cite{rho0CLAS}. It is supposed that in this case
a rise of $T_A$ as a function of $Q^2$ should indeed indicate an onset
of CT.

In Ref.~\cite{Frankfurt:2008pz} first theoretical estimates have been
made for the expected results. The framework used is essentially a
Glauber calculation, with the prehadronic interactions being described
by the pQCD-inspired cross section of Farrar et
al.~\cite{Farrar:1988me}. In this model the prehadronic cross section
consists of a $1/Q^2$-dependent starting value; the cross section then
increases linearly in time up to an expansion time where it reaches
the full hadron-nucleon cross section.  The result of
Ref.~\cite{Frankfurt:2008pz} was a transparency ratio $T_A$ strongly
rising with $Q^2$.

The main scope of Ref.~\cite{Frankfurt:2008pz} was a description of
FSI only.  The elementary reaction process played no role, \ie{}~all
the $\rho$ mesons produced experience CT, independent of their
production mechanism.  However, as we will demonstrate in this paper,
the knowledge of the elementary exclusive cross section $(e,e'\RHO{})$
off protons \emph{and} neutrons is a necessary prerequisite for a
proper description of the transparency ratio $T_A$ and thus a proof of
the CT effect.

In addition, in any analysis of CT it is essential to account for the
experimental acceptance limitations~\cite{Falter:2002vr}.  So far, the
acceptance conditions applied in the JLAB experiment and the interplay
of kinematical cuts with standard nuclear effects like Fermi motion
and FSI of $\rho^0$ and its $\pi^+\pi^-$ decay products have not been
studied.  We will show that the experimental cuts may strongly affect
the transparency ratio and produce at high values of $Q^2$ a behavior
which may overshadow and/or mimic the rise of $T_A$.

In the following we study the \RHO{} production on $^{12}$C and
$^{56}$Fe nuclei. We choose the kinematical conditions for our
calculations such that they correspond as nearly as possible to the
actual values of the ongoing experiment at JLAB~\cite{rho0CLAS}. In
particular we look for the region at smaller $Q^2$ where the
experimental values are centered and discuss also the event-types and
$\rho^0 \to \pi^+\pi^-$ reconstruction problems in this experiment.
The main focus of our studies is to clarify the observable
consequences of CT effect and investigate possible other effects that
might influense the wanted $\rho$-CT signal.  Such an analysis of CT
at present JLAB energies is challenging due to the combined
contribution of different effects such as the uncorrelated
$\pi^+\pi^-$ background affecting the $\rho^0\to \pi^+\pi^-$ signal,
the largely reduced (by experimental cuts) phase space, nuclear FSI
and Fermi motion effects.

The outline of the present paper is as follows. In Sec. II we describe
the model used in the calculation of exclusive $(e,e'\rho^0)$ cross
section of nucleons.  In Sec. III we introduce the GiBUU model used in
the calculations and also describe the hadronization and CT scenarios
followed in this work.  The experimental cuts used in the experiment
at JLAB are considered in Sec. IV. The effect of Fermi motion in the
transparency ratio is discussed in Sec. V. The results are presented
in Sec. VI. The conclusions are summarized in Sec. VII.

\begin{figure}[t]
  \begin{center}
    \includegraphics[width=1.\columnwidth,clip=true]{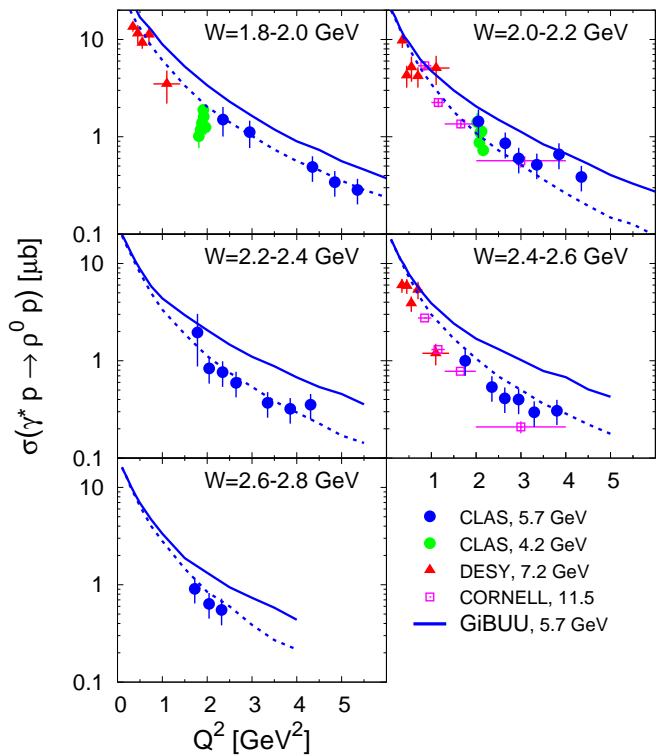}
    \caption{(Color online)
      $Q^2$ dependence of the integrated cross section in the
      reaction $p(\gamma^*,\rho^0)p$ (solid curves) for different $W$
      bins above the resonance region. The compilation of experimental
      data is taken from Ref.~\cite{Morrow:2008ek}.  The dashed curves
      describe the corresponding cross sections in the reaction
      $n(\gamma^*,\RHO)n$.  }
    \label{fig:Q2depRho}
  \end{center}
\end{figure}
\begin{figure}[t]
  \begin{center}
    \includegraphics[width=1.\columnwidth,clip=true]{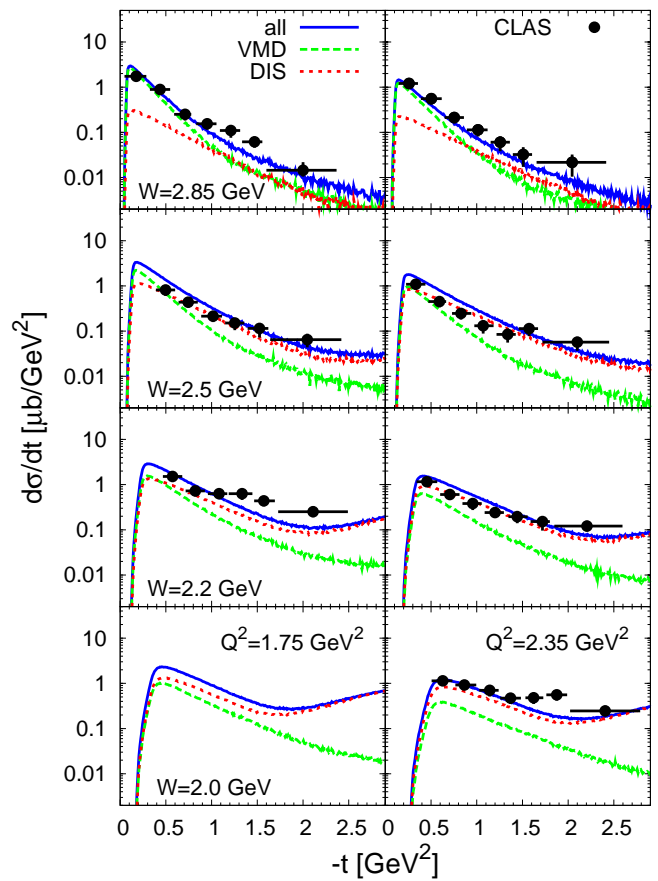}

    \caption{(Color online)
      Differential cross section $\dd\sigma/\dd t$ in the
      reaction $p(\gamma^*,\rho)p$ for different $(W,Q^2)$ bins. The
      experimental data are from Ref.~\cite{Morrow:2008ek}. The solid
      curves are the sum of VMD (dashed) and DIS (dotted)
      contributions.  }
    \label{fig:dsdt}
  \end{center}
\end{figure}
\begin{figure}[t]
  \begin{center}
    \includegraphics[width=1.\columnwidth,clip=true]{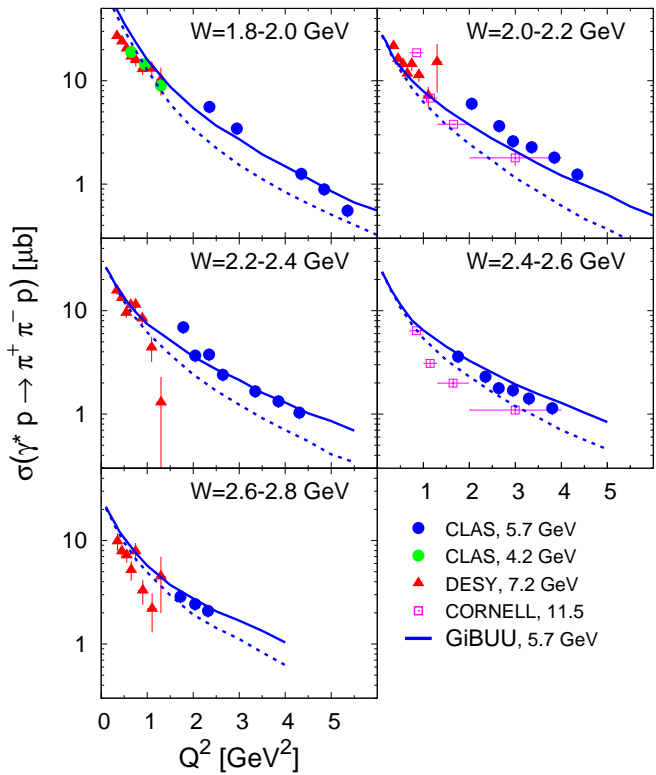}
    \caption{(Color online)
      $Q^2$ dependence of the integrated cross section in the
      reaction $p(\gamma^*,\pi^+\pi^-)p$ (solid curves) for different
      $W$ bins.  The compilation of experimental data is taken from
      Ref.~\cite{Morrow:2008ek}.  The dashed curves describe the
      corresponding cross sections in the reaction
      $n(\gamma^*,\pi^+\pi^-)n$.  } \vspace{-0.cm}
    \label{fig:pippim}
  \end{center}
\end{figure}
%

\section{\texorpdfstring{Exclusive $\rho^0(770)$ production}{Exclusive
    rho0(770) production}}\label{s:TR}

The exclusive production of mesons may involve very different
production mechanisms driven either by hadronic or by hard partonic
interactions. The CT occurs in the scattering of virtual photons off
partons.  Indeed, in Ref.~\cite{Kaskulov:2008ej} it was shown that the
exclusive $\pi^+$ electroproduction and the CT signal observed in this
reaction off nuclei~\cite{:2007gqa,Cosyn:2007er,Larson:2006ge} can be
understood if only the hard partonic
events~\cite{Kaskulov:2008xc,Kaskulov:2009gp}, that go through
high-lying excitations of the nucleon~\cite{Kaskulov:2010kf},
experience an expansion time and a corresponding CT effect.  In
analogy, the unambiguous identification of the $\rho$-CT effect
requires the understanding of the elementary \RHO{} production
mechanism. Also the elementary cross sections off nucleons are the
first input in any event generator attempting to describe the
corresponding reaction off nuclei.  We are interested in the
description of exclusive \RHO{} production around the values of
$W\simeq 1.8 \dots 2.8\GeV$. This is very difficult task because in
this region the contributions of large amount of nucleon excitations
should be accounted for explicitely, see Ref.~\cite{Obukhovsky:2009th}
and references therein.

To account for the contribution of nucleon resonances we consider the
DIS like interaction with partons since DIS involves the excitation of
all possible resonances.  To validate this assumption the high energy
event generator PYTHIA is used that -- besides diffractive vector
meson dominance (VMD) events -- also describes hard scattering events
in a string-fragmentation picture. In PYTHIA the exclusive \RHO{}
production is treated as exclusive limit, $z \to 1$, of semi-inclusive
DIS
\begin{equation}
  p(e,e'\rho^0)X
  \,{\longrightarrow}\,  p(e,e'\rho^0)p\quad \text{for}\quad z\to 1.
\end{equation}
This is in the spirit of the exclusive-inclusive
connection~\cite{Bjorken:1973gc}.

However, it is not obvious that the event generator designed for
inclusive reactions should also work in exclusive processes.  
In Figure~\ref{fig:Q2depRho} we, therefore, show the results for the $Q^2$
dependence of the integrated cross section for the exclusive reaction
$p(\gamma^*,\rho^0)p$ above the resonance region.  The $Q^2$ range
shown there is the one covered by the JLAB $\rho$-CT experiment.  The
curves correspond to the default parameters used in the PYTHIA
generator.

The Pomeron induced diffractive production of \RHO{} is isoscalar in
nature. On the contrary the DIS production of \RHO{} introduces an
isospin dependent component which makes the cross sections off protons
and neutrons unequal. Therefore, we also show in
Figure~\ref{fig:Q2depRho} the production of \RHO{} in the reaction
$n(\gamma^*,\rho^0)n$ off neutrons. The isospin dependence of the
\RHO{} production cross section and its contribution to $T_A$ have not
been considered before; the naive assumption $\sigma_p=\sigma_n$ was
always used. The latter has important consequences in isospin
asymmetric systems such as the $^{56}$Fe nucleus studied at JLAB. Note
that in the JLAB experiment the transparency ratio $T_A$ is normalized
to the cross section on the deuterium target.

In Figure~\ref{fig:dsdt} we show a comparison of the differential
cross section $\dd\sigma/\dd t$ in the exclusive reaction
$p(\gamma^*,\rho^0)p$ with the experimental data from
Ref.~\cite{Morrow:2008ek}; the contributions of DIS and diffractive
VMD events are given separately.  
The PYTHIA model describes the magnitude
and the shape of the differential spectra very well.

In Fig.~\ref{fig:dsdt} it can also be seen that in the region $W\simeq
2\GeV$ the DIS mechanism gives the dominant contribution whereas at
higher values of $W$ the diffractive VMD component dominates the
forward production of \RHO{}.  Concerning the $Q^2$ dependence the
region of low $Q^2$ is always dominated by the VMD events (see
discussion below). At large values of $Q^2$, however, the VMD
component is strongly decreasing and the \RHO{} production mechanism
in the forward and off-forward regions is driven by the partonic DIS
component.  Because in DIS the deep inelastic structure function is
factorized out of the fragmentation function the $Q^2$ dependence of
the transverse cross section in $p(e,e'\rho^0)p$ must essentially
follow the $Q^2$ dependence of the total DIS cross section. This
behavior has been already observed in exclusive $\pi^+$
electroproduction~\cite{Kaskulov:2008xc}. We find here that the
situation is similar in exclusive \RHO{} production and that the
$\rho^0$ experimental data follow this behavior as well.

Interestingly, the Lund Model used here for the
hadronization predicts two jets for the $\rho^0p$ final state in the
forward and backward directions. Therefore, there is a sizable
backward production of the \RHO{}.
In the experimental analysis of Ref.~\cite{Morrow:2008ek} the
off-forward region is extrapolated by assuming a functional form as
$e^{bt}$ thus missing the backward rise. Since, the backward
production of \RHO{} is large and has been not taken into account
in~\cite{Morrow:2008ek} the data may underestimate the actual
integrated cross section. This can be seen by comparing
Figure~\ref{fig:Q2depRho} and Figure~\ref{fig:dsdt} where the forward
angle data are consistent with the model while the integrated cross
sections are consistently lower than the model at higher values of
$Q^2$.

In the actual experiment the \RHO{} is reconstructed from its
$\pi^+\pi^-$ decay products. Therefore, one has to also control the
uncorrelated $\pi^+\pi^-$ background which may contaminate the
$\rho^0$ signal. The same string fragmentation picture is applied to
the uncorrelated $\pi^+\pi^-$ background as implemented in the PYTHIA
generator. In Figure~\ref{fig:pippim} we show the $Q^2$ dependence of
the integrated cross sections in the reaction $p(e,e'\pi^+\pi^-)p$ off
protons. The compilation of experimental data is from
Ref.~\cite{Morrow:2008ek}.  They include the correlated
$\rho\to\pi^+\pi^-$ as well as the uncorrelated $\pi^+\pi^-$ events.
The two basic mechanisms, diffractive production and hard DIS
production, together describe the data very well.

\begin{figure}[t]
  \begin{center}
    \includegraphics[width=1\columnwidth,clip=true]{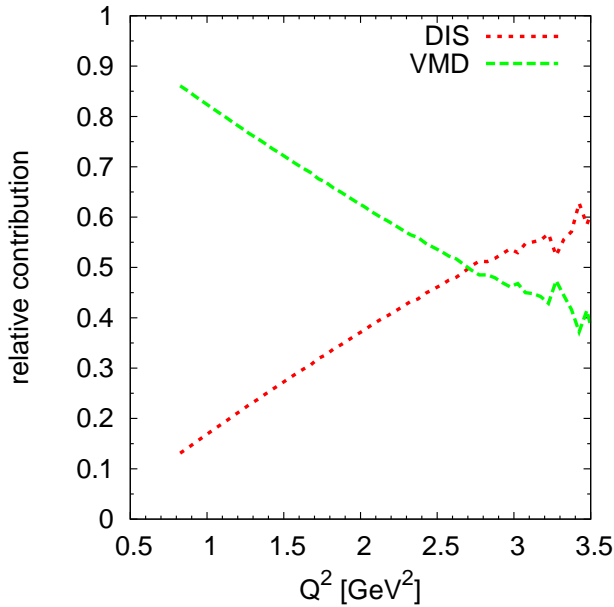}
    \caption{(Color online)
      The relative importance of VMD (green dashed) or DIS (red
      dotted) events compared to the total sample as function of $Q^2$
      for D target.  Only real \RHO{} are selected, there are no fake
      reconstructions.  } \vspace{-0.7cm}
    \label{fig:ratioVMDtotal}
  \end{center}
\end{figure}

Since the experimental cuts do not allow for a perfect separation of
diffractive events, we show in Figure~\ref{fig:ratioVMDtotal} the
relative importance of diffractive VMD and DIS events over the full
sample.  We show the result for deuterium D as an example, but the
conclusions are valid also for other nuclei, except for the
kinematical limitations at high $Q^2$. FSIs do not affect the relative
importance of the production mechanisms, but they do affect the
relative importance of the two components in the observed sample.

The relative importance of the two event types depends strongly on
whether we count all reconstructed \RHO{} or only those, which are
real \RHO{}. In DIS events the probability of reconstructing a \RHO{},
which is indeed a fake one, is much larger than in VMD events.  As one
can see in Figure~\ref{fig:ratioVMDtotal}, already in an experiment,
which can distinguish between 'true' and 'fake' \RHO{} mesons, the DIS
component is non-negligible over the full $Q^2$ range, increases
strongly with $Q^2$ and becomes even dominant for the values of $Q^2
\gtrsim 2.5\GeV^2$. For the actual experiment, where the fake and the
true \RHO{} mesons cannot be distinguished, this transition happens
even earlier (see Fig.~\ref{fig:dsdt}).

\section{\texorpdfstring{CT in the transport of $\rho^0$ in nuclei}{CT
    in the transport of rho0 in nuclei}}

The theoretical framework followed here is the GiBUU transport model
\cite{GiBUU}. In this model first the primary interaction of the
incoming electron with the nucleons of the target nucleus is described
within the impulse approximation; the latter assumes that the
interaction takes place with only one nucleon at a time and is
expected to be valid for the momentum transfers treated here.  Fermi
distribution of nucleons follows the local density approximation.

The fate of produced hadrons in FSI is then described by the
coupled-channel Boltzmann-Uehling-Uhlenbeck (BUU) transport equation
which describes the time evolution of the phase space density
$f_i(\vec r,\vec p,t)$ of particles of type $i$.  Besides the nucleons
these particles involve the baryonic resonances and mesons that can be
produced in FSI. For the baryons the equation contains a mean field
potential which depends on the particle position and momentum. The BUU
equations of each particle species $i$ are coupled via the mean field
and the collision integral. The latter allows for elastic and
inelastic rescattering and side-feeding through coupled-channel
effects; it accounts for the creation and annihilation of particles of
type $i$ in a secondary collisions as well as elastic scattering from
one position in phase space into another. The resulting system of
coupled differential--integral equations is solved via a test particle
ansatz for the phase space density. For fermions Pauli blocking is
taken into account via blocking factors in the collision term. The
model has been widely tested and validated for very different types of
reactions, from heavy-ion over hadron-induced reactions up to
electron- and neutrino-induced reactions on nuclei. All the details
about the GiBUU code and the comprehensive list of publications
describing the model can be found in~\cite{GiBUU}.

In electroproduction, the \RHO{} mesons are reconstructed from
$\pi^+\pi^-$ pairs, both in the actual experiment as well as in our
simulations.  Therefore, for realistic calculations, it is important
to account for the in-medium decay and attenuation of $\rho^0\to
\pi^+\pi^-$ decay pions.  These pions, after the $\rho^0$ decay inside
a nucleus, propagate further through the nucleus and experience FSI.
Thus, the final result, $T_A$, is not only affected by \RHO{}
absorption, but also by pion FSI. The most striking result of our
studies is the large effect of the pion interactions: half of the
observed attenuation is due to pion absorption.  Since the effect of
the pionic interactions is so large, it is not appropriate to
distinguish between absorption of the \RHO{} and some
'small'~\cite{Frankfurt:2008pz} pion absorption correction.  Instead
it is necessary to use a model, like the one used here, which is
capable of treating all the particle species and their interactions
consistently.

There is no definite framework to accommodate the notion of CT in
theoretical models. Different CT scenarios exist in the literature.
For example, in Ref.~\cite{Kopeliovich:1993pw,Kopeliovich:1993gk} a
color dipole model has been applied to high-energy reactions in which
all the hadronization happens outside the target nucleus. In the
present model, aimed at the description of experiments in a
kinematical regime where hadronization mostly happens inside the
nuclear target, we use the concept of {\it production} and {\it
  formation} times for production of a 'constituent' quark and
formation of a color-neutral hadron as developed in
\cite{Gallmeister:2005ad}.

The prehadronic interactions between the production time and the
formation time follow the pQCD-inspired quantum diffusion model of
Ref.~\cite{Farrar:1988me} assuming that the formation time corresponds
to the expansion time of a point-like configuration (PLC).  In this
picture the cross section in FSI grows linearly with time
$\tau$~\cite{Farrar:1988me}
\begin{eqnarray} \label{CTsigma}
  \label{SigmaQt}
  \sigma_{\rm eff}(Q^2,\tau)  = \sigma_{\rm tot} \hspace{6cm}  \\
  \times \left[\left( \frac{r_{\rm lead}}{Q^2}\left(1 - \frac{\tau}{\tau_f}\right)
      + \frac{\tau}{\tau_f}\right)\,\Theta(\tau_f - \tau) + \Theta(\tau - \tau_f)\right] \nonumber
\end{eqnarray}
where $r_{\rm lead}$ is the ratio of leading quarks to all quarks in
the hadron.  The scaling with $r_{\rm lead}$ guarantees that summing
over all particles in an event, on average the prefactor becomes
unity.  In the presence of the CT effect, see the factor $\sim 1/Q^2$
in Eq.~(\ref{SigmaQt}), the intranuclear attenuation of hadrons
propagating through the nuclear medium decrease as a function of
photon virtuality $Q^2$.

The hard part of the primary high energy electromagnetic interaction
is described by the Lund model which means that the final state
consists of an excited string. This string then fragments into
hadrons. An extraction of the {\it formation} time $\tau_f$ in the
target rest frame follows the space-time pattern of hadronization as
described in Ref.~\cite{Gallmeister:2005ad}.  This model has been
already successfully applied to the hadron attenuation experiments at
200-280\GeV{} (EMC), at 28\GeV{} (HERMES) and at 5\GeV{} (JLAB)
\cite{Gallmeister:2007an,Kaskulov:2008ej}.

In our earlier studies of the $\pi$-CT experiment
\cite{Kaskulov:2008ej} we had argued that only the DIS events should
experience CT.
For \RHO{} production the situation might be, however, different for
the diffractive process. Now the incoming photon can fluctuate into a
resolved \RHO{} component.  Experiments could thus be influenced also
by the prehadronic formation and expansion period of this
diffractively produced \RHO{} mesons.  It is, therefore, of interest
to study the relative importance of DIS to diffractive events. Such a
distinction could give valuable information on the mechanism of CT.
Therefore, in the following we will discuss three different scenarios
for the transparency ratio $T_A$ and CT :
\begin{enumerate}
\item no CT present; all hadrons in FSI interact with their full
  hadronic cross section from their creation vertex on
\item CT is only present in the hard partonic events which are
  determined by the underlying fragmentation model
\item both the hard partonic and soft diffractive events experience CT
  effect.
\end{enumerate}

Throughout the following we shall also compare three different
calculations:
\begin{enumerate}
\item calculations for a deuterium target including relative momenta
  of proton and neutron, including final state interactions (FSI)
  (albeit being a very small effect), are labeled by ``D''
\item calculations for a nucleus, including Fermi motion and Pauli
  blocking, but no FSI (except particle decays) will be denoted by
  ``A0'' as e.g. ``Fe0'' for an Iron target
\item finally calculations on Iron as above, but now including the
  full FSI machinery are labeled by ``Fe'' (or in general by ``A'').
\end{enumerate}

The calculations for finite nuclei are done here without the shadowing
effect, expected to still be quite small at the energies and momenta
relevant here. Since the coherence length has been kept nearly
constant in the JLAB experiment the shadowing corrections are not
effective in the $Q^2$ dependence of $T_A$.

\section{Experimental Cuts} \label{s:expcuts}

Within the GiBUU transport model, one simulates the production of
\RHO{} mesons (among others) on a Monte Carlo basis. All produced
particles, independent of their production mechanism, are propagated
through the nucleus according the transport equations. At the end we
have four-vectors of all final state particles. This enables us to
simulate all the experimental reconstruction of $\pi^+\pi^-$ events
and cutting.

\begin{figure}[t]
  \begin{center}
    \includegraphics[width=1.1\columnwidth,clip=true]{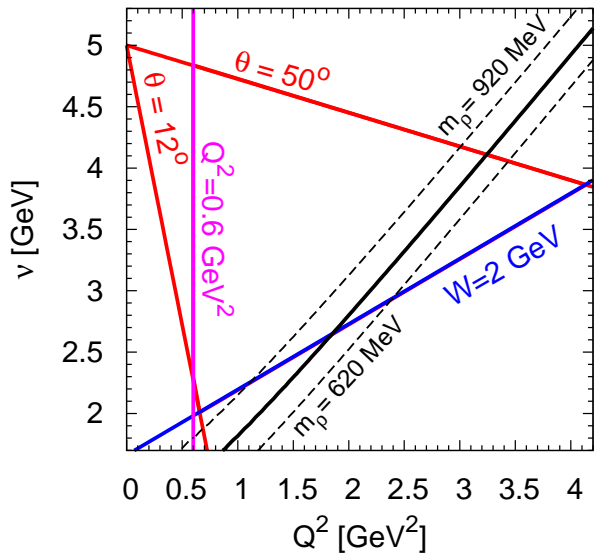}
    \caption{(Color online)
      Experimental coverage in the $\nu$--$Q^2$--plane
      corresponding to the electron beam energy $E_e=5\GeV$.  Lines
      correspond to the various cuts ($W>2\GeV$, $Q^2>0.6\GeV^2$,
      $\theta_e>12^o$ and $\theta_e<50^o$).  The upward sloping lines
      labeled by $m_\rho$ indicate the cut at $t_0=-0.4\GeV^2$ (solid)
      according eq.~(\ref{eq:t01}) with $m_\rho=0.770\GeV$ and
      $m_\rho=0.770\pm0.150\GeV$ (dashed curves). The rightmost curve
      corresponds to the smallest mass value.  With this cut the
      region of large $Q^2$, to the right of the thick straight line,
      is forbidden for on-shell $\rho$ production on a nucleon at
      rest.}
    \label{fig:t0_cut}
  \end{center}
\end{figure}

We start our considerations of acceptance conditions by looking at
kinematics.  In Figure~\ref{fig:t0_cut} we show the accessible
kinematical region in the $\nu$-$Q^2$ plane for an electron beam
energy $E_e=5\GeV$.  As can be seen there, the understanding of the
acceptance in the region of small electron scattering angles is of
major importance, since it influences strongly the possible minimal
values of $\nu$ or $W$ in the region of $Q^2 \lesssim 1\GeV^2$.  Since
effects of Fermi motion have most impact at low $W$, the variation in
minimal $W$ translates into the question, whether one observes them or
not.  In the following we will only use the acceptance
cuts~\cite{HafidiPriv} for the scattered electron and assume full
detection efficiency within these cuts ($\theta_e=12^o\dots50^o$,
$Q^2>0.6\GeV^2$).

In addition, the JLAB experiment~\cite{rho0CLAS} applies the cuts for
the invariant mass $W$, the momentum transfer $t$ and the incoming
photon energy $\nu$.
\begin{enumerate}
\item $W>2\GeV$ in order to avoid the resonance region,
\item $t>-0.4\GeV^2$ to be in the diffractive region,
\item $t<-0.1\GeV^2$ to exclude coherent production off the nucleus,
\item $z = E_\rho/\nu>0.9$ to select the elastic process; here
  $E_\rho$ is the energy of the \RHO{} meson produced.
\end{enumerate}

We note here, that the first cut relies on the assumption of a
quasi-free process since the invariant mass $W$ is calculated for a
nucleon at rest, neglecting any Fermi motion. The actual minimal
values for $W$ for the elementary reaction to be considered in the
calculations may, due to Fermi motion, reach much smaller values.
Thus the first cut does not guarantee the desired region above the
resonances.

The understanding of the second cut is of major importance for the
interpretation of the high $Q^2$ part of the considered experiment.
To see how the $t > -0.4 \GeV^2$ cut acts in the ($Q^2,\nu$) plane we
now consider the kinematical limits for $t$ in a collision on a
nucleon at rest with $W^2 = M^2_N -Q^2 +2M_N\nu$. These are given by
($t_1\leq t\leq t_0$)
\begin{eqnarray}
  \label{eq:t01}
  t_{0,1}=\left(\frac{-Q^2-m_\rho^2}{2W}\right)^2 - \hspace{5cm} \\
  \left(\frac{
      2M_N\sqrt{\nu^2+Q^2}\mp\sqrt{(W^2+m_\rho^2-M^2_N)^2-4W^2m_\rho^2}}{2W}\right)^2. \nonumber
\end{eqnarray}
%
where the notations for the variables are obvious.  These purely
kinematical limits for a nucleon at rest (or the deuterium target)
directly translate into expressions for $(Q^2,\nu)$ at fixed
$t_{0,1}$.  The bound $t_0$, together with the experimental cut
$t>-0.4\GeV^2$, translates also into a relation between $\nu$ and
$Q^2$ and limits the values of $Q^2$ at fixed $\nu$.  This cut is
shown in Figure~\ref{fig:t0_cut} by the three parallel upward sloping
lines. For the region of large $Q^2$ to the right of the lines on a
nucleon (or deuterium) target no \RHO{} production with an on-shell
mass is possible. However, the spectral function of the \RHO{} meson
smears out this sharp cut. The two thinner dashed lines give a measure
for the effects of the \RHO{} mass distribution.

For heavier nuclei all these considerations about the $t$ cuts are
smeared out further by Fermi motion. As a consequence, the cut is not
so effective for such nuclei. We have indeed tested that with a
realistic Fermi momentum distribution the cut on $t_0$ has no
consequences for heavy nuclei, while it has a major effect for
deuterium.  As a consequence, any ratio of nuclear and nucleon
(deuterium) cross sections will become very large in the large $Q^2>
2.5\GeV^2$ region since there the denominator gets very small. This
will be demonstrated in the following Section~\ref{sec:FermiMotion}.

Within our model we can not calculate coherent \RHO{} production off
nuclei. We, therefore, apply the third cut as it is used in the data
analysis.  The fourth cut, however, is again accessible within our
model. The exclusivity cut on high energies of the produced $\rho$ is
meant to enrich the $\rho$-CT signal and we apply it to our results.

As for the detection of the scattered electron, we will also assume
full detection efficiency for the pions.

\begin{figure}[b]
  \begin{center}
    \includegraphics[width=0.9\columnwidth,clip=true]{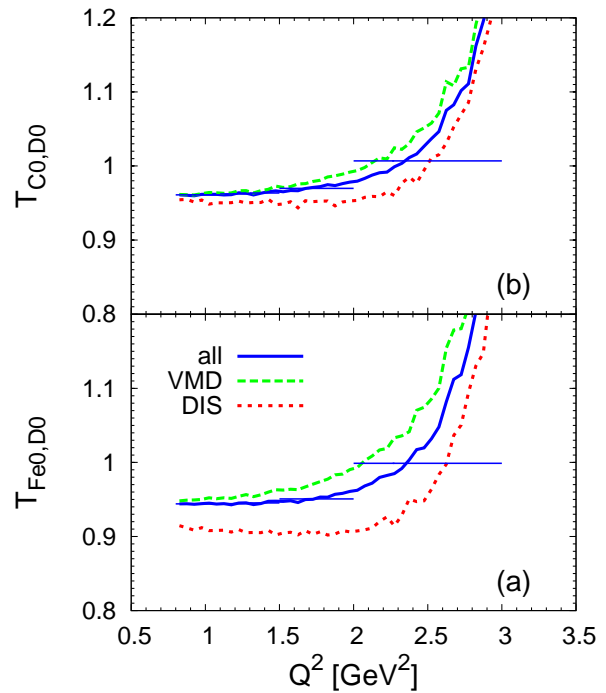}
    \caption{(Color online)
      The transparency ratio for nuclei without FSI, \ie{}~the
      effect Fermi motion, for Fe (a) and C target (b).  Only
      true $\rho$ are used.  The contributions of the processes are
      indicated by line style as 'all' (blue solid), 'VMD' (green
      dashed) and 'DIS' (red dotted). The thin horizontal lines repeat
      the results for 'all', but now with a coarse binning in $Q^2$.}
    \label{fig:FermiMotion} \vspace{-0.7cm}
  \end{center}
\end{figure}
\begin{figure*}[t]
  \begin{center}
    \includegraphics[width=2.\columnwidth,clip=true]{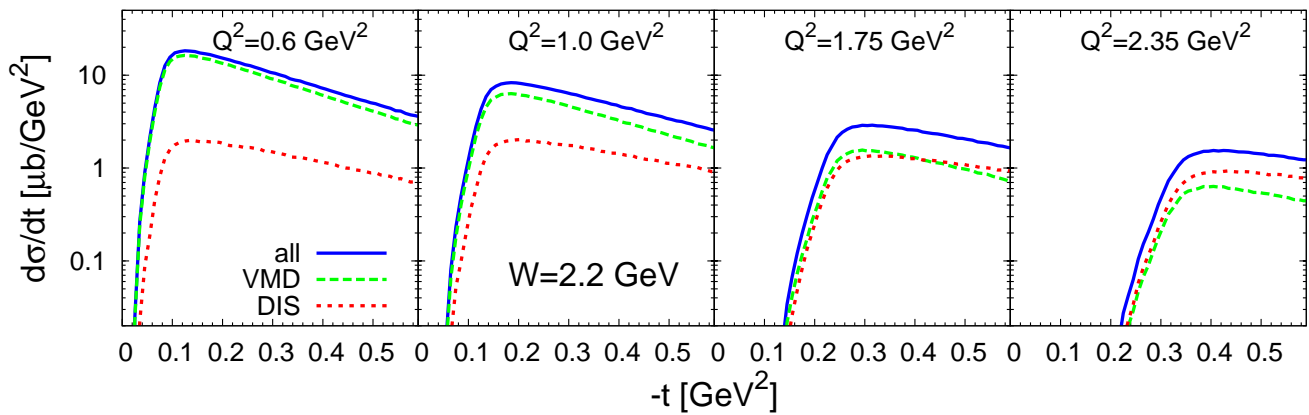}
    \caption{(Color online)
      Differential cross section $\dd\sigma/\dd t$ in the
      reaction $p(\gamma^*,\rho)p$ for different values of $Q^2$ and a
      value of $W=2.2\GeV$ in the kinematics of JLAB experiment.  The
      solid curves are the sum of VMD (dashed) and DIS (dotted)
      contributions.  }\vspace{-0.7cm}
    \label{fig:dsdtJLAB}
  \end{center}
\end{figure*}
%

\section{Effect of Fermi Motion}\label{sec:FermiMotion}

At first we demonstrate a trivial effect which can mimic the $\rho$-CT
signal in the transparency ratio $T_A$. This effect is not generic and
is merely tied to the special set of $t$ kinematical cuts used in the
actual JLAB experiment.

Since common to all hadronization models considered in the following
we discuss first the effect of Fermi motion alone on the transparency
ratio by analyzing the simplest ratio ``A0/D0'', \ie{}~the ratio of
nuclear cross sections for some nucleus (without FSI) over
calculations for deuterium (without FSI). In the final $\pi^+\pi^-$
Monte Carlo sample all the experimental cuts discussed above are taken
into account.  As one can see in Figure~\ref{fig:FermiMotion}, for
both $^{12}$C/D and $^{56}$Fe/D ratios we observe that this
transparency ratio rises sharply to values above 1 for $Q^2$ values
larger than $\simeq 2.5\GeV^2$ following -- at smaller $Q^2$ -- that
of the VMD component, which is dominant here. This increase is a
consequence of Fermi motion alone, since all FSI are turned off. This
artificial rise of $T_A$ for $Q^2 > 2.5 \GeV^2$ is due to the $t$-cut
discussed in Section~\ref{s:expcuts} that influences the denominator
(the cross section off the deuterium target) in the transparency ratio
much more strongly than the numerator.  In
Figure~\ref{fig:FermiMotion} we also show the results obtained when a
coarse binning in $Q^2$ similar to that in the experiment is applied.
In this case the rise of the curves is not so visible anymore.
However, a bin expanding from $Q^2=2\GeV^2$ to $Q^2=3\GeV^2$ is still
clearly affected.

Thus, Fermi motion alone can mimic a behavior with $Q^2$ that is
qualitatively expected for CT. At lower values of $Q^2$ the rise of
$T_A$ shows up only in the VMD and not in the DIS component. This is
due to the fact that the partonic DIS events are more isotropically
distributed, whereas the diffractive events are forward peaked. Then
the $t$-cut is not so effective.

The forward differential cross sections $\dd\sigma/\dd t$ at
$W=2.2\GeV$ and for different values of $Q^2$ bins are shown in
Figure~\ref{fig:dsdtJLAB}. Different slopes of different \RHO{}
production components can be clearly seen.  In the kinematics of the
JLAB experiment the low $Q^2$ region is dominated by the diffractive
VMD component. On the contrary, the high $Q^2$ region is partonic.

Figure~\ref{fig:FermiMotion} also shows that $T_A < 1$ for $Q^2 <
2.5\GeV{}^2$. This lowering of $T_A$ even in the absence of any FSI is
again due the experimental $t$-cut described above.

Note that, the effects of Fermi motion on the transparency could be
minimized by normalizing the transparency ratio $T_A$ to the $^{12}$C
cross section.

\begin{figure}[b]
  \begin{center}
    \includegraphics[width=0.9\columnwidth,clip=true]{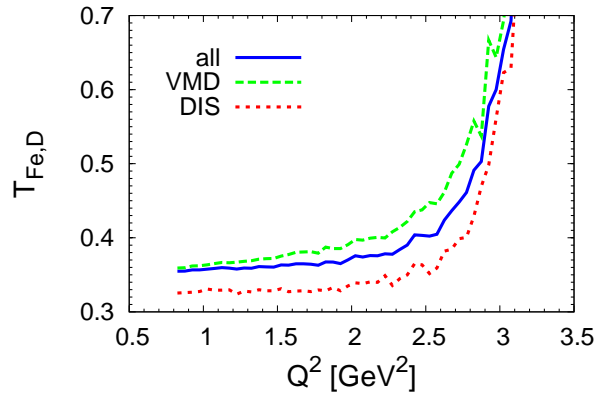}
    \caption{(Color online)
      The transparency ratio of \RHO{} for Fe target in the
      scenario without CT. The contributions of the processes are
      indicated by line style as 'all' (blue solid), 'VMD' (green
      dashed) and 'DIS' (red dotted).  }
    \label{fig:N_FeD}
  \end{center}
\end{figure}
%

\section{Results on Transparency}

\subsection{No CT}

The transparency for the model in which all created hadrons interact
with their full hadronic cross section from their point of creation on
is shown in Figure~\ref{fig:N_FeD}.  It is seen that the transparency
-- as in the case with Fermi motion only -- increases only weakly
until $Q^2 \approx 2 \GeV^2$ and then rises steeply for larger $Q^2$.
This latter strong rise is -- as already discussed -- due to the
$t$-cut and is present even for the case of Fermi motion alone, see
Figure~\ref{fig:FermiMotion}.

It is worthwhile to separate our results into their origins and to
look at the question, whether VMD or DIS induced \RHO{} are attenuated
differently. Therefore we show in Figure~\ref{fig:N_FeD} the results
separately for DIS and VMD \RHO{}s.  While both the VMD and the DIS
parts noticeably increase with $Q^2$ the resulting total curve 
 is flatter than the VMD
contribution because with increasing $Q^2$ also the weight of DIS
events increases. Since these DIS events are more strongly attenuated
the transparency $T_A$ is being held down until the $t$-cut effects
prevail.

\subsection{CT only for DIS events}

We turn now to the discussion of the CT only for hard DIS events.  In
the previous considerations we have assumed that all particles
interact immediately after their creation with their full cross
section, \ie{}~that there are no CT effects at work. In earlier work
we have shown that this assumption does not hold for higher beam
energies \cite{Gallmeister:2007an} or for exclusive pion production at
CLAS energies\cite{Kaskulov:2008ej}. Therefore we now use a
hadronization picture as developed in
\cite{Gallmeister:2005ad,Gallmeister:2007an}. In this picture the
hadronic interactions cross section in FSI grows linearly with
time~\cite{Farrar:1988me}, see Eq.~(\ref{SigmaQt}).  The expansion
time $\tau_f$, which we identify with the formation time in the
definition of \cite{Gallmeister:2005ad}, is obtained from the string
breaking mechanism as outlined in \cite{Gallmeister:2005ad}.  On the
other hand, the \RHO{} from diffractive VMD events are assumed to
start interacting with their full hadronic cross section immediately
after their production.

\begin{figure}[t]
  \begin{center}
    \includegraphics[width=0.9\columnwidth,clip=true]{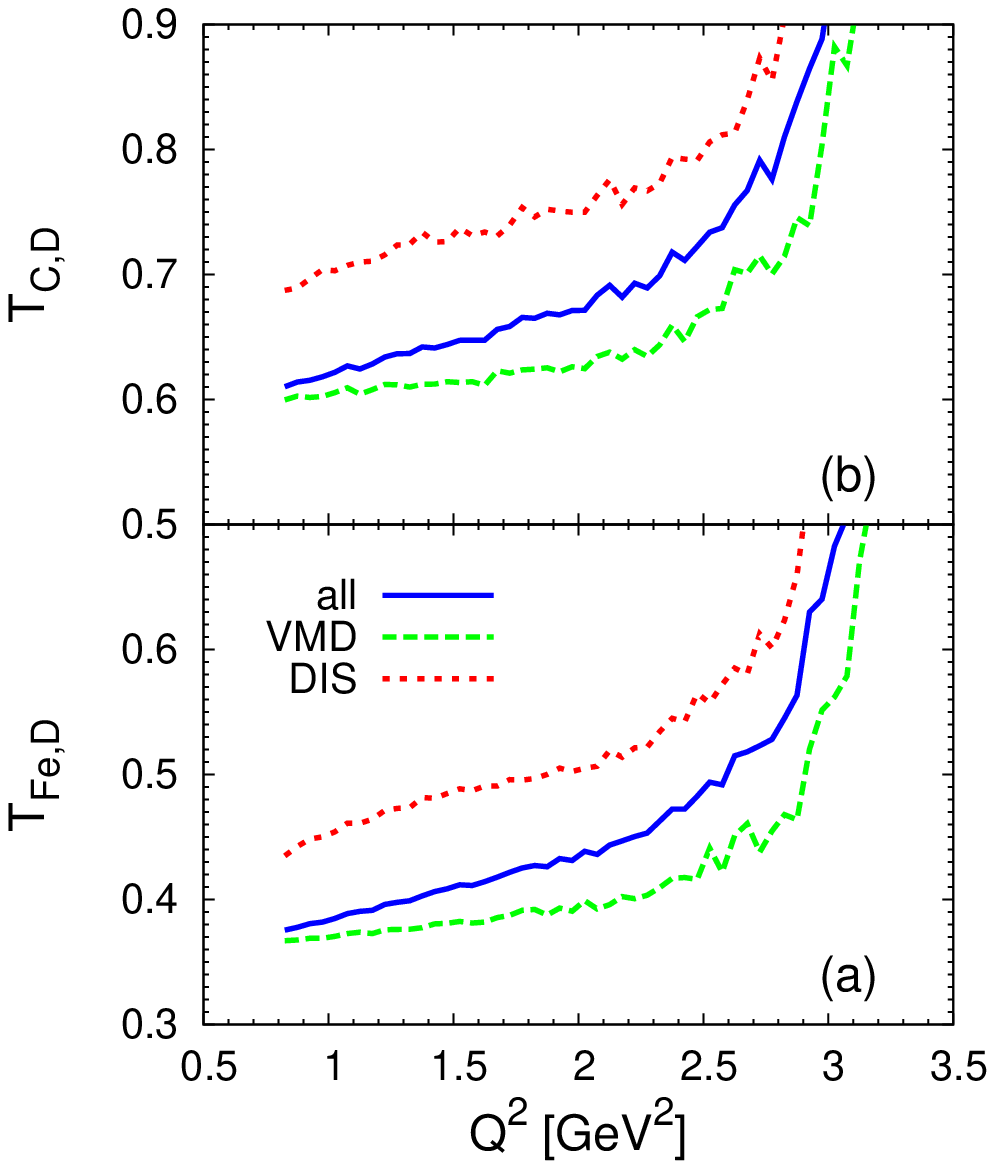}
    \caption{(Color online)
      The transparency ratio of \RHO{} for Fe (a) and C (b)
      target in the scenario, where only the DIS part has CT
      effects.  The contributions of the processes are indicated by
      line style as 'all' (blue solid), 'VMD' (green dashed) and 'DIS'
      (red dotted).  } \vspace{-0.2cm}
    \label{fig:N_CFeD}
  \end{center}
\end{figure}
\begin{figure}[t]
  \begin{center}
    \includegraphics[width=0.9\columnwidth,clip=true]{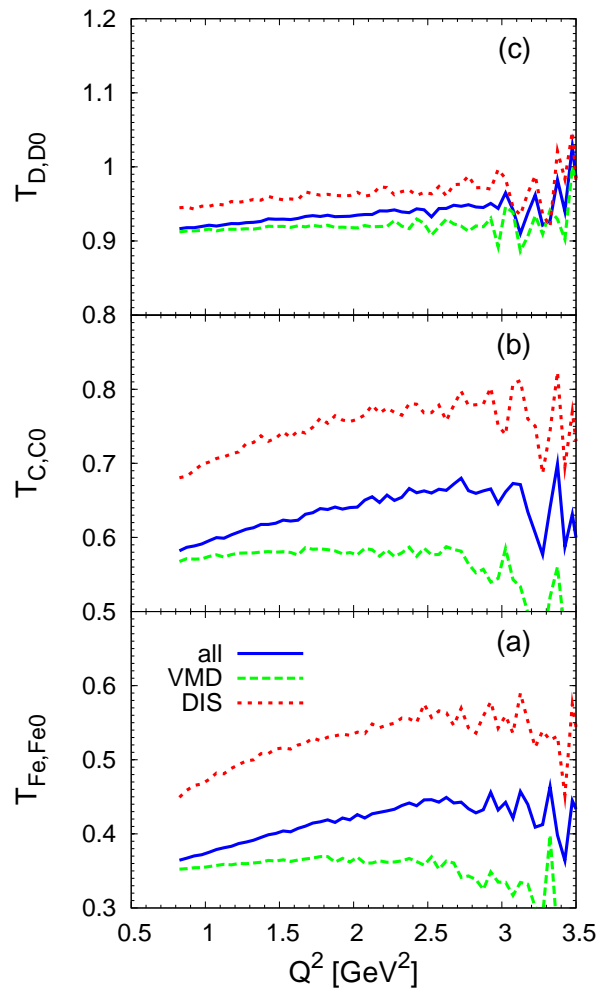}
    \caption{(Color online)
      The ratio of all \RHO{}s produced on $^{56}$Fe (a),
      $^{12}$C (b) and D (c) to that on
      the same targets, respectively, but without FSI.  The
      contributions of the processes are indicated by line style as
      'all' (blue solid), 'VMD' (green dashed) and 'DIS' (red dotted).
    }
    \label{fig:FeFe0}
  \end{center}
\end{figure}

In Figure~\ref{fig:N_CFeD} we show the result on the hadron
attenuation.  The rise of the VMD component is the same as that in
Fig.~\ref{fig:N_FeD} and is entirely caused by the Fermi motion. It
dominates $T_A$ at small $Q^2$. At large $Q^2$ we find a stronger
increase with $Q^2$ compared to the case without CT, driven by the
weakened interaction of the DIS-like events during the expansion time.
The \RHO{}s stemming from DIS events are less attenuated, while the
\RHO{} from VMD events are unaffected (as expected). Since in the
final ratio we have an admixture of both origins such that the weight
of DIS events increases with $Q^2$ also the overall transparency
increases with $Q^2$ in this scenario. In the scenario without CT
effect, see Figure~\ref{fig:N_FeD}, the partonic DIS contribution is
more strongly attenuated than the VMD part and there the mixing has an
opposite effect.

In order to separate the effects of CT from those originating in the
Fermi motion we show in Figure~\ref{fig:FeFe0} the ratio of the \RHO{}
production on D, $^{12}$C and $^{56}$Fe to that on the same targets,
respectively, without any FSI.  The rise of the DIS component of the
transparency in Fig.~\ref{fig:FeFe0} with $Q^2$ is due to the
$1/Q^2$-dependence of the first term in Eq.~\ref{CTsigma}. We note,
however, that already for a deuterium target, the FSI introduce an
effect in the order of 5-10\%.

 \subsection{CT for DIS and VMD events}

In a next step, in order to study the CT effects also for VMD events,
we assume in a third model that all \RHO{} have a finite formation
time $t_F=\gamma \tau_F$ in the laboratory. Here $\gamma$ is the
Lorentz boost factor, while the formation time in the rest frame of
the \RHO{} is taken to be $\tau_F=0.4\fm$, the value used in
\cite{Frankfurt:2008pz}\footnote{The string-expansion times extracted
  from PYTHIA, which were used in scenario 2, are on average larger by
  about a factor of 2. This observation is important because the (not
  well-known) expansion time has a crucial influence on absolute
  height and the slope of $T_A$ as a function of $Q^2$.}

\begin{figure}[t]
  \begin{center}
    \includegraphics[width=0.9\columnwidth,clip=true]{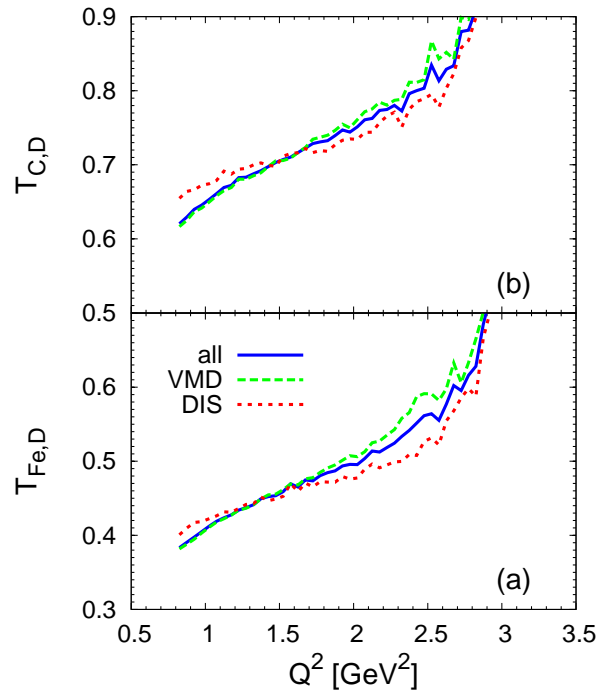}
    \caption{(Color online)
      The transparency ratio of \RHO{} for $^{56}$Fe (a)
      and $^{12}$C (b) target in a scenario where all the
      produced $\rho$ experience the CT effect. The contributions of
      the processes are indicated by line style as 'all' (blue solid),
      'VMD' (green dashed) and 'DIS' (red dotted).  } 
    \label{fig:N_CFeD2}
  \end{center}
\end{figure}

In this last extreme picture, see Figure~\ref{fig:N_CFeD2}, one
observes a strong increase of the attenuation ratio with $Q^2$. It is
essentially the same for both VMD and DIS components.

The final results in Fig.~\ref{fig:N_CFeD2} show a very steep rise of
$T_A$ with $Q^2$ which seems to be ruled out by the preliminary
experimental data. However, before jumping to such conclusions we
recall that the final result depends on a number of ingredients, which
all conspire with each other in affecting the observable transparency.
Among them is -- foremost -- the Fermi-motion and its interplay with
the experimental $t$-cuts. The influence of the latter also varies
with the $t$-dependence of the experimental cross section.

We also note that a large part of the observed \RHO{} suppression is
due to pion reabsorption in FSI. This then raises interesting, but
hard to answer questions on the \RHO{} decay during its expansion time
and on the inclusion of CT effects for pions from \RHO{} decay. The
pion FSI then complicate any experimental determination of CT for the
$\rho$ meson.

%

\section{Conclusions}

In this paper we have investigated the effects of various experimental
(cuts) and theoretical (Fermi-motion) properties on an experiment
aiming for a verification of color transparency at JLAB energies.  We
have illustrated that a careful analysis of experimental acceptances
and cuts can have an influence on the final results because these cuts
tend to include different kinematical regions in the numerator and the
denominator of the transparency ratio. As a consequence, the
elementary production cross sections do not drop out in the
transparency ratio, but have to be explicitly taken into account. The
observed transparency is then no longer a measure for FSI alone, but
is also affected by all these effects just mentioned. This has
consequences, for example, for the interpretation of a recent
experiment at JLAB on nearly-exclusive \RHO{} production.

We have also shown that in this experiment different event types (VMD
and DIS) contribute significantly in different $Q^2$ ranges and -- due
to their different absorption -- can affect the behavior of the
transparency.  Furthermore, Fermi motion can smear the $W$ and $t$
ranges and, as a consequence, leads to a strong rise of the
transparency with $Q^2$ at large momentum transfers. This latter
behavior is always there and has to be separated from the rise of
$T_A$ with $Q^2$ expected from CT.

Our calculations show that in the range of smaller values of $Q^2 <
2.2 \GeV^2$ CT indeed leads to a larger rise than that given by Fermi
motion alone. The rise -- above that expected from Fermi motion alone
-- is directly proportional to the expansion time. It could thus be
possible to determine the latter from a careful analysis of the data.

A complementary experimental proposal for the future JLAB facilities
would be to measure the transparency ratio and $\rho$-CT in the decay
of \RHO{} into dilepton $e^+e^-$ pairs, thus continuing the successful
g7 experiment that worked with real photons \cite{Wood:2008ee}.  This
could provide much cleaner sample of events which are not contaminated
by the in-medium FSI of pions as in the hadronic $\rho\to \pi^+\pi^-$
reconstruction experiments.

\begin{acknowledgments}
  The authors thank Kawtar Hafidi and Lamia El Fassi for their
  patience explaining to us their detector and experimental
  procedures.  We gratefully acknowledge helpful discussions with the
  whole GiBUU group.  We also gratefully acknowledge support by the
  Frankfurt Center for Scientific Computing, where parts of the
  calculations were performed. This work was supported by the HIC for
  FAIR, by BMBF and by DFG under SFB/TR16.
\end{acknowledgments}




\begin{thebibliography}{99}

\bibitem{Brodsky:1988xz} S.~J.~Brodsky and A.~H.~Mueller,
  Phys.\ Lett.\ B {\bf 206}, 685 (1988)


\bibitem{Frankfurt:1992dx} L.~Frankfurt, G.~A.~Miller and M.~Strikman,
  Comments Nucl.\ Part.\ Phys.\ {\bf 21}, 1 (1992).

\bibitem{Jain:1995dd} P.~Jain, B.~Pire and J.~P.~Ralston,
  Phys.\ Rept.\ {\bf 271}, 67 (1996).

\bibitem{Jennings:1991rw} B.~K. Jennings and G.~A. Miller, \newblock
  Phys. Lett. {\bf B274}, 442 (1992).

\bibitem{Farrar:1988me} G.~R. Farrar, H.~Liu, L.~L. Frankfurt and
  M.~I. Strikman, \newblock Phys. Rev. Lett. {\bf 61}, 686 (1988).

\bibitem{Airapetian:2002eh} A.~Airapetian {\it et
    al.}, 
  Phys.\ Rev.\ Lett.\ {\bf 90}, 052501 (2003).


\bibitem{Aclander:2004zm} J.~L.~S.~Aclander {\it et al.},
  Phys.\ Rev.\ C {\bf 70}, 015208 (2004).

\bibitem{Carroll:1988rp} A. S. Carroll {\it et al.}, Phys. Rev. Lett.
  {\bf 61}, 1698 (1988).

\bibitem{Adams:1994bw} M.~R.~Adams {\it et
    al.}, 
  Phys.\ Rev.\ Lett.\ {\bf 74}, 1525 (1995).

\bibitem{Falter:2002vr} T.~Falter, K.~Gallmeister and U.~Mosel,
  \newblock Phys. Rev. {\bf C67}, 054606 (2003).

\bibitem{Kopeliovich:2001xj} B.~Z. Kopeliovich, J.~Nemchik, A.~Schafer
  and A.~V. Tarasov, \newblock Phys. Rev. {\bf C65}, 035201 (2002).

\bibitem{rho0CLAS} CLAS, K.~Hafidi {\em et~al.}, \newblock in
  preparation.

\bibitem{Frankfurt:2008pz} L.~Frankfurt, G.~A. Miller and M.~Strikman,
  \newblock Phys. Rev. {\bf C78}, 015208 (2008).

\bibitem{Kaskulov:2008ej} M.~M.~Kaskulov, K.~Gallmeister and U.~Mosel,
  Phys.\ Rev.\ C {\bf 79}, 015207 (2009).

\bibitem{:2007gqa} B.~Clasie {\it et al.},
  Phys.\ Rev.\ Lett.\ {\bf 99}, 242502 (2007).

\bibitem{Cosyn:2007er} W.~Cosyn, M.~C.~Martinez and J.~Ryckebusch,
  Phys.\ Rev.\ C {\bf 77}, 034602 (2008).

\bibitem{Larson:2006ge} A.~Larson, G.~A.~Miller and M.~Strikman,
  Phys.\ Rev.\ C {\bf 74}, 018201 (2006).

\bibitem{Kaskulov:2008xc} M.~M.~Kaskulov, K.~Gallmeister and U.~Mosel,
  Phys.\ Rev.\ D {\bf 78}, 114022 (2008).

\bibitem{Kaskulov:2009gp} M.~M.~Kaskulov and U.~Mosel,
  Phys.\ Rev.\ C {\bf 80}, 028202 (2009).

\bibitem{Kaskulov:2010kf} M.~M.~Kaskulov and U.~Mosel,
  Phys.\ Rev.\ C {\bf 81}, 045202 (2010).

\bibitem{Obukhovsky:2009th} I.~T.~Obukhovsky {\it et
    al.},
  Phys.\ Rev.\ D {\bf 81}, 013007 (2010).

\bibitem{Bjorken:1973gc} J.~D.~Bjorken and J.~Kogut,
  Phys.\ Rev.\ D {\bf 8}, 1341 (1973).

\bibitem{Morrow:2008ek} S.~A.~Morrow {\it et al.}  [CLAS
  Collaboration],
  Eur.\ Phys.\ J.\ A {\bf 39}, 5 (2009)

\bibitem{GiBUU} GiBUU homepage, \newblock
  http://gibuu.physik.uni-giessen.de

\bibitem{Kopeliovich:1993pw}
  B.~Z.~Kopeliovich, J.~Nemchick, N.~N.~Nikolaev and B.~G.~Zakharov,
  Phys.\ Lett.\  B {\bf 324}, 469 (1994).

\bibitem{Kopeliovich:1993gk}
  B.~Z.~Kopeliovich, J.~Nemchick, N.~N.~Nikolaev and B.~G.~Zakharov,
  Phys.\ Lett.\  B {\bf 309}, 179 (1993)
  [arXiv:hep-ph/9305225].

\bibitem{Gallmeister:2005ad} K.~Gallmeister and T.~Falter, \newblock
  Phys. Lett. {\bf B630}, 40 (2005).

\bibitem{Gallmeister:2007an} K.~Gallmeister and U.~Mosel, \newblock
  Nucl. Phys. {\bf A801}, 68 (2008).

\bibitem{HafidiPriv} K.~Hafidi, L.~El Fassi, \newblock private
  communication.

\bibitem{Wood:2008ee} M.~H.~Wood {\it et al.} [CLAS Collaboration],
  Phys.\ Rev.\ {\bf C78}, 015201 (2008).



\end{thebibliography}
\end{document}